\documentclass[aps,prb,reprint,showpacs,amsmath,amssymb,superscriptaddress,citeautoscript]{revtex4-1}
\usepackage{graphicx}    
\usepackage{color}

\begin{document}


\title{Manifestation of the Berry curvature in Co$_{2}$TiSn Heusler films }

\author{Benedikt Ernst}
\affiliation{Max Planck Institute for Chemical Physics of Solids, N\"{o}thnitzer Str. 40, D-01187 Dresden, Germany}%

\author{Roshnee Sahoo}\email{sahoo@cpfs.mpg.de}
\affiliation{Max Planck Institute for Chemical Physics of Solids, N\"{o}thnitzer Str. 40, D-01187 Dresden, Germany}%

\author{Yan Sun}
\affiliation{Max Planck Institute for Chemical Physics of Solids, N\"{o}thnitzer Str. 40, D-01187 Dresden, Germany}%

\author{Jayita Nayak}
\affiliation{Max Planck Institute for Chemical Physics of Solids, N\"{o}thnitzer Str. 40, D-01187 Dresden, Germany}%

\author{Lukas M\"{u}chler}
\affiliation{Max Planck Institute for Chemical Physics of Solids, N\"{o}thnitzer Str. 40, D-01187 Dresden, Germany}%

\author{Ajaya K. Nayak}
\affiliation{National Institute of Science Education and Research, Jatni, Bhubaneswar-752050, India}%

\author{Nitesh Kumar}
\affiliation{Max Planck Institute for Chemical Physics of Solids, N\"{o}thnitzer Str. 40, D-01187 Dresden, Germany}%

\author{Anastasios Markou}
\affiliation{Max Planck Institute for Chemical Physics of Solids, N\"{o}thnitzer Str. 40, D-01187 Dresden, Germany}%

\author{Gerhard H. Fecher}
\affiliation{Max Planck Institute for Chemical Physics of Solids, N\"{o}thnitzer Str. 40, D-01187 Dresden, Germany}%

\author{Claudia Felser}
\affiliation{Max Planck Institute for Chemical Physics of Solids, N\"{o}thnitzer Str. 40, D-01187 Dresden, Germany}%

\date{\today} 

\begin{abstract}

Various Co$_{2}$ based Heusler compounds are predicted to be Weyl materials.
These systems with broken symmetry possess a large Berry curvature, and
introduce exotic transport properties. The present study on epitaxially grown
Co$_{2}$TiSn films is an initial approach to understand and explore this
possibility. The anomalous Hall effect in the well-ordered Co$_{2}$TiSn films
has been investigated both experimentally and theoretically. The measured Hall
conductivity is in good agreement to the calculated Berry curvature. Small
deviations between them are due to the influence of skew scattering on the Hall
effect. From theoretical point of view, the main contribution to the anomalous
Hall effect originates from slightly gapped nodal lines, due to a symmetry
reduction induced by the magnetization. It has been found that only part of the
nodal lines contributed near to the anomalous Hall conductivity at a fixed Fermi
energy which can be explained from a magnetic symmetry analysis. Furthermore,
from hard x-ray photoelectron spectroscopy measurements, we establish the
electronic structure in the film that is comparable to the theoretical density
of states calculations. The present results provide deeper insight into
the spintronics from the prospect of topology.

\end{abstract}

\maketitle

\section{INTRODUCTION}

Recently, spintronics has fascinated many researchers as it has abundant
advantages over conventional electronics~\cite{Parkin1,Pulizzi2}. Heusler alloys
have attracted much interest in this regard due to their high Curie temperature
($T_{C}$) and easy tunability in terms of structural, electronic or magnetic
properties~\cite{FeFe2013,Graf6,Kubler4,Kandpal5,Sahoo7}. The Co$_2$-based
Heusler alloys have captivated certain attentions, as many of them are predicted
to be half-metallic ferromagnets with full spin polarization at the Fermi
energy~\cite{Kandpal5,Kubler8,Galanakis9}. Few of these materials have already
been successfully fabricated in to multilayered structures exhibiting giant
magnetoresistance (GMR) and tunnel magnetoresistance (TMR) for nonvolatile
memory applications~\cite{Sato10,Tezuka11}. In a junction consisting of
Co$_{2}$FeMnSi based Heusler alloys a GMR of 74.8\% has been reported by Sato
{\it et al.}~\cite{Sato10}. Similarly, a large TMR ratio of 386\% has been
achieved at room temperature using Co$_{2}$FeAlSi~\cite{Tezuka11}. Recently, a
remarkable TMR ratio of 2000\% at 4.2~K (354\% at 300~K) was achieved using
epitaxial Co$_{2}$MnSi magnetic tunnel junctions~\cite{Liu12}. The Co$_{2}$TiSn
compound is also one of the promising candidates that may be beneficial for spin
manipulation due to its high Curie temperature and half metallic
properties~\cite{Meinert13,Barth14}. Bulk Co$_{2}$TiSn has been studied vividly
in terms of electronic and magnetic properties~\cite{Barth14,Kandpal15,Barth16}.
For spintronics application as well as for fundamental understanding, it is
essential to implement such systems into thin films.

In the present work, we discuss the role of the Berry curvature on the transport
properties of Co$_{2}$TiSn. Addressing the physics of the Berry curvature is
intriguing  as it helps to understand intrinsic or dissipationless Hall currents
in the systems that are useful for spintronic applications. Many theoretical and
experimental studies have been performed to understand the anomalous Hall effect
using the Berry curvature approach~\cite{Nagaosa17}. In recent studies, large
anomalous Hall effects (AHE) have been reported in the non-collinear
antiferromagnets Mn$_{3}$Ge and Mn$_{3}$Sn~\cite{Nayak18,Nakatsuji19}. Such a
large anomalous Hall effect in materials with net zero magnetic moment is
understood in terms of a non-vanishing Berry
curvature~\cite{Nayak18,Nakatsuji19}. Recently, Co$_{2}$ based Heusler compounds
have been predicted to be Weyl materials~\cite{Nakatsuji19,Kubler20}. The
symmetry breaking in these materials bring fascinating transport properties due
to large Berry phase~\cite{Nakatsuji19,Kubler20,Wang21,Kubler22}. In the present
manuscript we focus on Co$_{2}$TiSn and analyze the above possibility with help
of Berry curvature calculations.

\section{THEORETICAL AND EXPERIMENTAL DETAILS}

We have calculated the electronic band structure by using the Vienna ab-initio
simulation package (VASP)~\cite{Kresse23,Kresse24}. The exchange correlation
functional was considered in the generalized gradient
approximation~\cite{Perdew25}. In all calculations the experimental lattice
parameters has been used. The magnetization was set along (001). In order to
calculate the intrinsic anomalous Hall conductivity, we have projected the Bloch
wave functions into maximally localized Wannier
functions~\cite{Marzari26,Souza27,Mostofi28}. Based on the tight binding model
Hamiltonian, the anomalous Hall conductivity was calculated by the Kubo formula
approach at the clean limit~\cite{Nagaosa17}:

\begin{equation}
\begin{aligned}
\sigma_{xy}^{z} = ie^{2}\hbar \left ( \frac{1}{2\pi} \right )^{3} \int d\vec{k}\underset{E(n,\vec{k})<E_{F}}{\sum}f(n,\vec{k})\Omega_{n,xy}^{z}(\vec{k}), \\
\Omega_{n,xy}^{z}(\vec{k})= 2Im\underset{n'\neq n}{\sum} \frac{\langle n,\vec{k}|\hat{v}_{x}|n',\vec{k}\rangle \langle n',\vec{k}|\hat{v}_{y}|n,\vec{k}\rangle}{(E(n,\vec{k})-E(n',\vec{k}))^{2}},
\end{aligned}
\end{equation}

where $f(n,\vec{k})$ is the Fermi-Dirac distribution, $E(n,\vec{k})$ is the
eigenvalue of the $n^{\rm th}$ eigenstate of $|u(n,\vec{k})>$ at $\vec{k}$
points, and ${v_{\alpha}}=\frac{1}{\hbar}\frac{\partial {H(\vec{k})}}{\partial
k_{\alpha}}$ is the velocity operator. A $500 \times 500 \times 500$ grid of $k$
points was used in the evaluation of the integrals.

Thin films of Co$_{2}$TiSn were grown on polished MgO(001) substrates by 
co-sputtering from elementary Co, Ti, and Sn targets. These films are named
according to their growth temperatures such as CTS-550, CTS-600 and CTS-650,
grown at $550^{\circ}$C, $600^{\circ}$C and $650^{\circ}$C, respectively. The
CTS-650 film was annealed for 30~minutes at $450^{\circ}$C and named as 
CTS-650+A. All films were capped with 2-3~nm of aluminum at room temperature to
prevent from oxidation. The stoichiometry of the films was checked with energy
dispersive x-ray spectroscopy (EDX). The crystal structure and thickness of the
films were determined using a standard Pananalytical x-ray diffractometer using
Cu-$K_{\alpha}$ radiation. Magnetic properties were measured in a Quantum Design
magnetometer (MPMS 3). The transport properties were performed in a Quantum
Design physical properties measurement system (PPMS). Hard x-ray photoelectron
spectroscopy (HAXPES) on the Co$_2$TiSn thin films was performed at beamline
BL47XU of Spring-8 (Japan).  The electron energy distribution was recorded by a
hemispherical energy analyser (R4000-HV, Scienta). The photon energy was fixed
at $h\nu=7.94$~keV using a double crystal monochromator (Si(111)) with post
monochromator (Si(444)). The overall energy resolution including monochromator
and analyzer resolution was determined by fitting the Au Fermi edge and was
found to be 250~meV at 300~K.  The experiment was carried out in near normal
emission geometry (see References~\onlinecite{OFF2015,FF2013} for more details of the
HAXPES experiments).

\section{RESULTS AND DISCUSSIONS}

\subsection{Berry curvature calculation}

Co$_{2}$ based Heusler compounds are proposed to be Weyl materials, where, the
band crossing points have been termed as "magnetic
monopoles"~\cite{Kubler20,Wang21}. These band crossing points near the Fermi
energy may lead to exotic transport phenomena~\cite{Kubler20,Wang21}. Here, we
have calculated the Berry curvature for Co$_{2}$TiSn as presented in Figure~1.
Co$_{2}$TiSn exhibits the face centered cubic L$2_{1}$ structure. In absence of
any net magnetic moment, Co$_{2}$TiSn possesses three mirror planes $M_{x}$,
$M_{y}$, and $M_{z}$, which protects the gapless nodal line like band structure
in the $k_{x}$=0, $k_{y}$=0 and $k_{z}$=0 planes,
respectively~\cite{Wang21,Chang29}. If the magnetic moment is considered, the
symmetry is reduced. For example, as performed in our experiments, if we apply a
magnetic field along the $z$ direction then the mirrors $M_{x}$ and $M_{y}$ are
not symmetry planes any more, while $M_{z}$ is still a symmetry plane since the
$z$ component of the spin $S_{z}$ is left invariant by $M_{z}$. Therefore, the
gapless nodal line only exists in the $k_{z}=0$ plane, and hence, the nodal
lines in the $k_{x}=0$ and $k_{y}=0$ planes exhibit a finite band gap due the
magnetic moment oriented in the $z$ direction. Additionally, two Weyl points
with a topological charge of $\pm2$ emerge along $k_{z}$ that derives from the
nodal line~\cite{Wang21,Chang29}. The intrinsic contribution of Weyl Fermions to
the anomalous Hall conductivity (AHC) is proportional to their $k$ space
distance along the magnetic field direction. However, we do not observe a direct
contribution of these points from the band structure calculations. This might be
explained by the fact that the Weyl points are several hundred meV above the
Fermi energy and due to the presence of other Weyl points in planes of constant
$k_{z}$~\cite{Chang29}.

\begin{figure*}[htb]
\centering
\includegraphics[angle=0,width=16cm,clip]{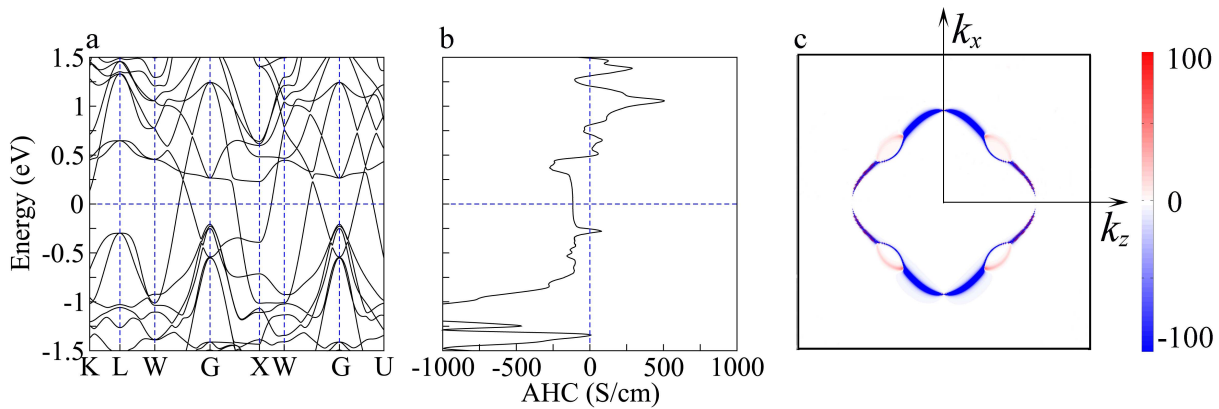}
\caption{\label{FIG 1}(Color online) 
         (a) Electronic band structure along high symmetry lines. 
         (b) Energy dependent anomalous Hall conductivity ($\sigma_{xy}^{z}$). 
         (c) Berry curvature ($\Omega_{xy}^{z}$) distribution in $k_{y}=0$ plane. The color bar is in arbitrary unit.}
\end{figure*}

The gapless nodal line band structure contains non zero Berry curvatures around
it, however, they are helically distributed in the mirror plane and the total
flux is zero. Hence the gapless nodal line in $k_{z}=0$  plane does not
contribute to the intrinsic anomalous Hall conductivity. On the other hand, the
nodal lines in $k_{x}=0$  and $k_{y}=0$  planes are broken by the magnetization,
which also force the Berry curvature to orient along the direction of
magnetization. Taking $k_{y}=0$  plane as an example, as presented in Figure~1c,
the $k$ points around the broken nodal lines are dominated by negative
$\Omega_{xy}^{z}$  components of the Berry curvature, and directly contribute to
the negative anomalous Hall effect, which shares the same mechanism of strong
spin Hall effect in nodal line materials~\cite{Sun30,Sun31}. Since the nodal
lines in mirror planes have strong energy dispersion~\cite{Wang21,Chang29}, only
part of the nodal lines contribute to the anomalous Hall conductivity at fixed
Fermi level. The intrinsic anomalous Hall conductivity almost keeps constant at
about 100~S/cm in the range of ~-0.25 to ~0.25~eV, which is just inside the
energy window of the dispersion of the nodal lines, see Figure~1a and b.
Therefore, the anomalous Hall conductivity in Co$_{2}$TiSn mainly originates
from the magnetization induced gaps by breaking the nodal line band structure.

\subsection{Structural properties of the thin films}

\begin{figure}[htb]
\centering
\includegraphics[angle=0,width=8.5cm,clip]{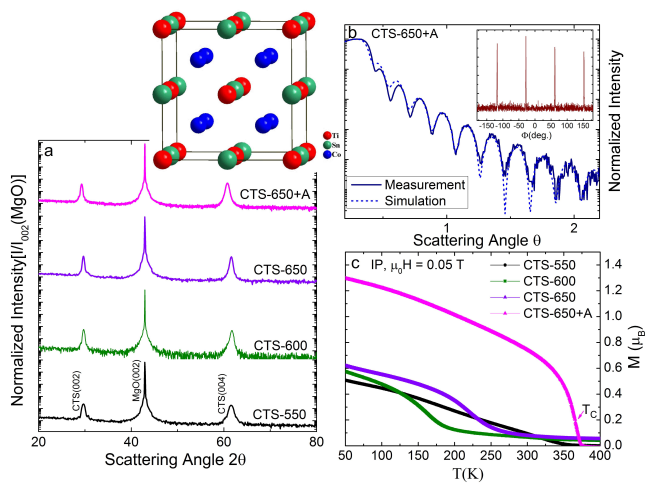}
\caption{\label{FIG 2}(Color online) 
         (a) $\theta-2\theta$ XRD diffraction data recorded at room temperature for different Co$_{2}$TiSn films. 
             The inset shows the $L2_1$ cubic crystal structure of the present films. 
         (b) XRR data for CTS-650+A. The inset shows azimuthal ($\phi$) scans along (111) plane. 
         (c) Temperature dependence of the in-plane (IP) magnetization curves for Co$_{2}$TiSn films measured at 0.05~T.}
\end{figure}

In this and the following sections we focus on the experimental findings,
starting with the crystal structure of the epitaxial Co$_{2}$TiSn films grown
under different conditions as mentioned in the experimental section. Figure~2a
depicts $\theta-2\theta$ XRD pattern for various Co$_{2}$TiSn films. All films
exhibit (002) and (004) peaks of the cubic $L2_1$ Heusler structure along with
(002) peak of the MgO substrate. The $L2_1$ crystal structure of Heusler
compounds is presented in the inset of Figure~2a. In this structure, Co , Ti and
Sn atoms occupy the Wyckoff positions 8c, 4b, and 4a, respectively.

The cubic lattice parameter CTS-550 is found to be $a=6.02$~{\AA}. The shift of
the XRD peaks towards lower $2\theta$ value points on an increase of the lattice
parameter with higher growth temperatures. The lattice parameter for CTS-650+A
is finally 6.1~{\AA}. The epitaxial relationship of the films with substrate is
found to be [100]~Co$_{2}$TiSn ${\parallel}$ [110]~MgO. X-ray reflectivity (XRR)
was measured to determine the thickness and surface smoothness of the film. The
XRR measurement for CTS-650+A is shown in Figure~2b. The clear oscillation of
fringes suggests that the surface of this film is smooth and homogeneous. Upon
fitting the XRR curve, the thickness and roughness are estimated to be 22~nm and
0.3~nm, respectively. The $\phi$-scan performed along the (111) plane displays
four sharp maxima at equal intervals, exemplifying four fold symmetry of the
present $L2_1$ cubic structure (inset in Figure~2b).

From EDX analysis the composition of CoTiSn films is confirmed to be 2:1:1 within error of 5\%.

\subsection{Hard x-ray photoelectron spectroscopy}

Besides by XRD and XRR measurements, the films were further investigated with
respect to their quality by Hard x-ray photoelectron spectroscopy. HAXPES is a
powerful technique that is utilized to understand the detailed electronic
structure of all type of materials~\cite{Zegenhagen34}. It helps to investigate
the bulk electronic structure as was explained in detail for Heusler compounds
in References~[\onlinecite{Fecher35,Ouardi36}]. The high kinetic energy
electrons exhibit a large probing depth inside the material and provide with
bulk properties that are free of surface effects. In particular, the details of
the valence band density of states are probed due to the integration over a
large volume in momentum space at high kinetic energies. Therefore, the HAXPES
valence band spectrum (see Figure~3b) has been measured for the well-ordered
CTS-650+A and is compared to the calculated density of states as shown in
Figure~3a. In the valence band spectrum the high intensity at -9.3~eV arises
from the contribution of $s$ electrons that are mainly localized close to the Sn
atoms according to the calculations. On the other hand, a minimum appears at
about -7.3~eV that reflects the $sp$ hybridization gap in the density of states
and is typical for Heusler compounds. In the upper part of the spectrum, four
maxima emerge at -5.8, -3.5, -2.0, and -0.4~eV binding energies. The feature at
-3.5~eV arises from the $pd$ hybridization of Co $d$ states with Sn $p$ states.
Closer to the Fermi energy, the spectra are dominated by $d$ states localized
mainly at the Co atoms. The valence band spectrum thus reflects the most
dominant features of the calculated density of states, accounting for matrix
elements that redistribute the intensities compared to the densities.

\begin{figure}[htb]
\centering
\includegraphics[angle=0,width=8.5cm,clip]{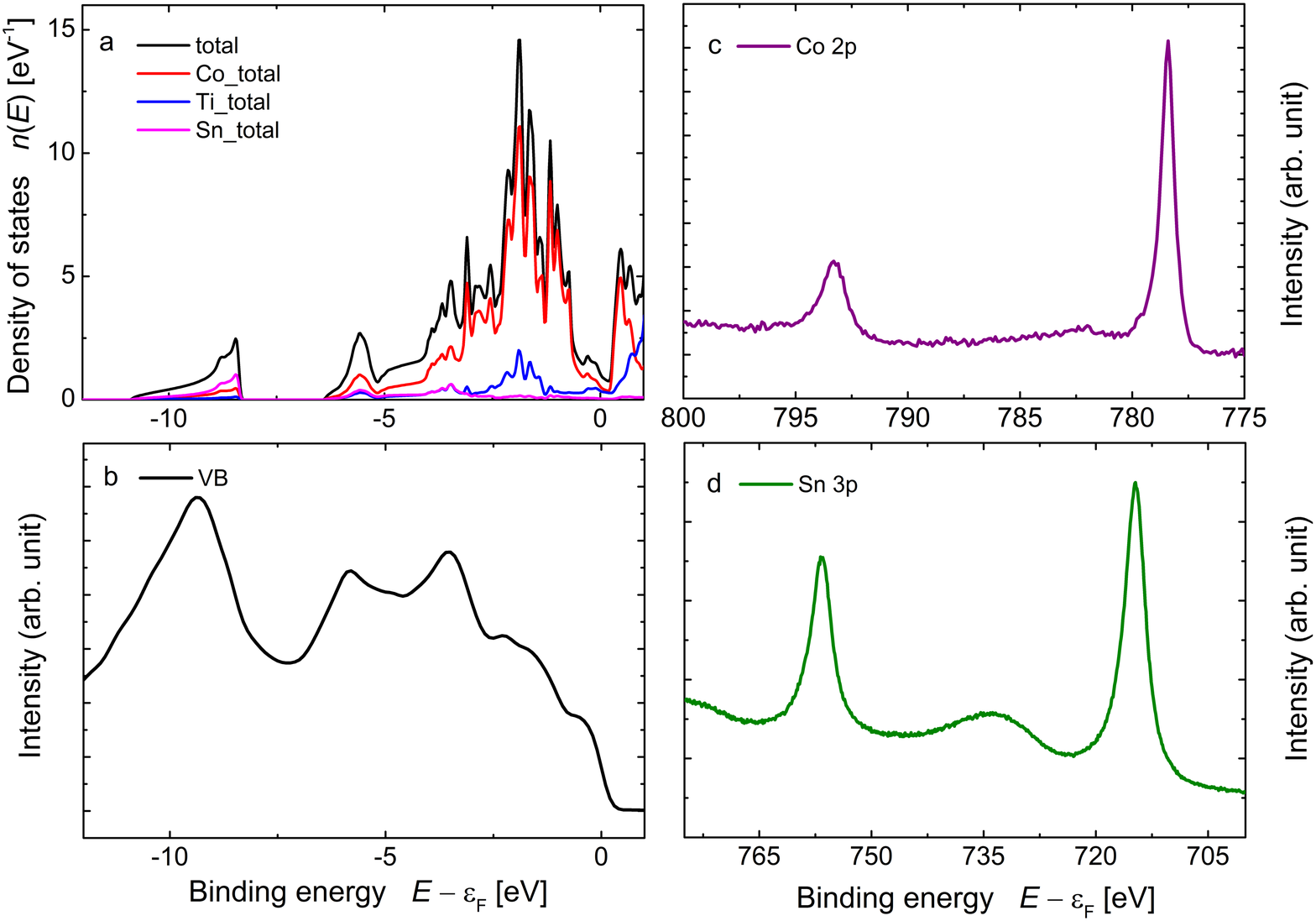}
\caption{\label{FIG 3}(Color online) 
         (a) Theoretical density of states (DOS) calculation for Co$_{2}$TiSn film. 
         (b) Hard x-ray valence band spectra, 
         (c) Co $2p$ and (d) Sn $3p$ core-level spectra of CTS-650+A measured by employing hard x-rays.}
\end{figure}

Core level spectra are used to explain the cleanness of the compound and with
restrictions its electronic state, chemical order, and composition. The Co $2p$
core level spectrum (Figure~3c) exhibits a distinct spin-orbit splitting of
14.9~eV. A similar size of the spin-orbit splitting (14.95~eV) was also reported
for polycrystalline samples~\cite{Barth14}. No oxide related features at 2-3~eV
higher binding energy compared to the Co main peak are detected. The Sn $3p$ core
level spectra exhibit a large spin orbit splitting of about 42~eV (Figure~3d)
together with pronounced side maxima. These broad transitions emerge not as
primary excitations but as energy loss structures from secondary band--band or
plasmon excitations. The later result from the free electron type behavior of
the $s$ electrons localized mostly close to Sn, as mentioned above. Again, no
oxygen or other impurity related features are detected. This indicates, together
with the Co $2p$ spectra, the cleanness of the sample. In all cases, no
significant chemical shifts of core level binding energies are found in
Co$_{2}$TiSn. These signify the metallic character of the compound.  All
observed features are in qualitative agreement with the previous HAXPES
measurement on polycrystalline bulk samples~\cite{Barth14}, reflecting the high
quality of the thin films.

\subsection{Magnetic and electrical properties}

This section reports on the magnetic properties and Hall effect measurements of
the epitaxially grown Co$_{2}$TiSn thin films. Figure~2c shows the temperature
dependent magnetization curves $M(T)$ that are measured in field cooled mode in
presence of an in-plane field of 0.05~T. For CTS-550, the magnetic transition is
quite broad in nature without a well-defined Curie temperature ($T_C$). With
increasing growth temperatures CTS-600 and CTS-650 exhibit a well-defined $T_C$
of about 160~K and 225~K, respectively. However, the non-zero magnetization
above $T_C$ for these films indicates the presence of some disorder. We found by
annealing of CTS-650 that a very sharp $T_C$ of $\approx366$~K is obtained for
the CTS-650+A film. For bulk Co$_{2}$TiSn, a $T_C$ of 355~K has been reported by
Barth {\it et al.}~\cite{Barth14}. It should be mentioned here that the trend of
$T_C$ in the observed films is an important factor to determine the degree of
order~/~disorder in the system. The observation of undefined or low $T_C$ in
CTS-550, CTS-600, and CTS-650 suggests the presence of large chemical disorder.
For instance, Meinert {\it et al.}~\cite{Meinert13} have reported a considerable
decrease in atomic disorder with increasing deposition temperature. The presence
of $D0_3$ type Co-Ti antisite disorder, as found by NMR and M{\"o}{\ss}bauer
measurements, has also been reported for bulk~\cite{Kandpal5}. Therefore,
disorder plays an important role in the present system. It can be reduced
considerably using higher substrate temperatures during growth.

\begin{figure}[htb]
\centering
\includegraphics[angle=0,width=8.5cm,clip]{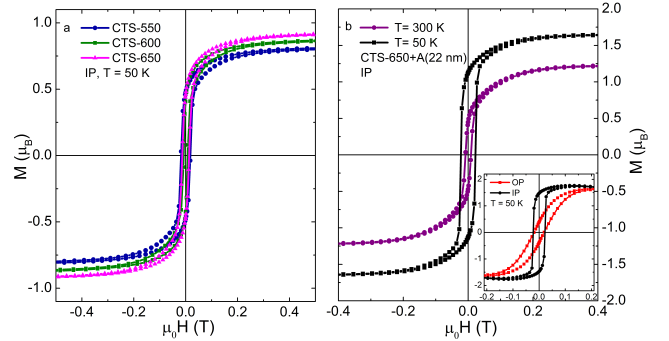}
\caption{\label{FIG 4}(Color online) Isothermal magnetization curves of 
          (a) CTS-550, CTS-600 and CTS-650 at 50~K, 
          (b) CTS-650+A at 50~K and 300~K. 
          The Inset shows in plane (IP) and out of plane (OP) curves of CTS-650+A at 50~K.}
\end{figure}

In-plane magnetic hysteresis loops $M(H)$ for different Co$_{2}$TiSn films
measured at 50~K are presented in Figure~4a. The saturation magnetic moment per
formula unit increases with increasing growth temperature from 0.8~$\mu_B$ for
CTS-550 to 0.9~$\mu_B$  for the CTS-650. For CTS-650+A the saturated moment
further increases to 1.6~$\mu_B$  at 50~K and 1.2~$\mu_B$ at 300~K (Figure~4b).
The presence of an in-plane easy axis is confirmed by the measurement of out of
plane (OP) $M(H)$ loops that exhibit a hard-axis type hysteresis (Figure~4b) in
the present film. The finding of a larger magnetic moment and higher $T_C$ for
CTS-650+A suggest that the improvement of the chemical order by annealing helps
to enhance the magnetic properties.

\begin{figure}[htb]
\centering
\includegraphics[angle=0,width=8.5cm,clip]{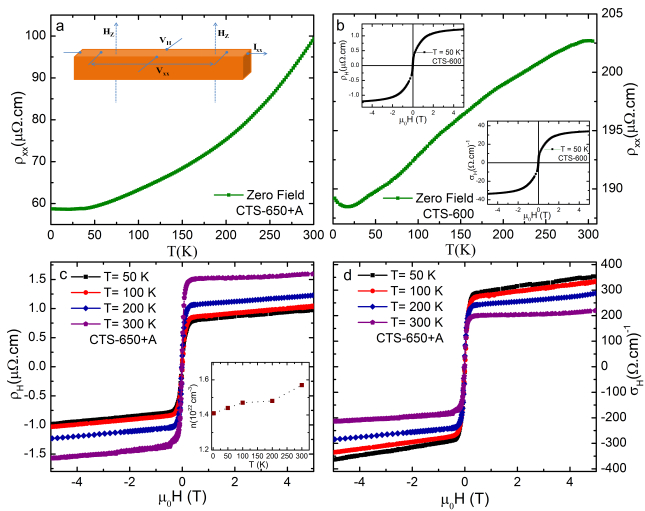}
\caption{\label{FIG 5}(Color online) Temperature dependence of linear electrical resistivity ($\rho_{xx}$) 
         (a) for CTS-650+A and 
         (b) for CTS-600. 
             The inset of (a) shows a schematic diagram for the measurement of linear voltage ($V_{xx}$) and lateral voltage ($V_H$).
             The inset of (b) shows the field dependence of Hall resistivity and Hall conductivity at 50~K for CTS-600. 
         (c) Field dependence of Hall resistivity at different temperatures for CTS-650+A. 
             The inset shows the temperature dependent carrier concentration. 
         (d) Field dependent Hall conductivity at various temperatures for CTS-650+A.}
\end{figure}

To characterize the AHE in the present thin films, we have carried out detailed
electrical transport measurements. A schematic diagram of the present
measurements is presented in the inset of Figure~5a. In the given configuration,
the current ($I_{xx}$) is applied along [100], the magnetic field ($H_z$) is
applied along [001] and the lateral Hall voltage ($V_H$) is measured along
[010]. As shown in Figure~5a, the temperature dependence of the linear
resistivity ($\rho_{xx}$) measured at zero field exhibits a metallic
characteristic. $\rho_{xx}$ values of 58~$\mu\Omega$cm and 100~$\mu\Omega$cm are
obtained at temperatures of 2~K and 300~K, respectively. Although CTS-600
displays a metallic behavior in the resistivity measurement, the magnitude of
resistivity is fairly higher than that of CTS-650+A. The higher residual
resistivity of CTS-600 clearly indicates the presence of higher disorder, as
discussed earlier. The residual resistivity ratio is calculated from the
resistivities at low (2~K) and high (300~K) temperatures:
RRR~=~$\rho_{xx}$(300 K)/$\rho_{xx}$(2 K). 
For CTS-650+A, we find a RRR value of 1.7, whereas, it is 1.07 for CTS-600. The
lower RRR value implies higher impurity scattering due to larger disordered in
CTS-600.

The Hall resistivity $\rho_H$ is obtained by measuring the transverse Hall
voltage $V_H$ as shown in the schematic diagram in the inset of Figure~5a. In a
ferromagnet, $\rho_H$ is usually written as,

\begin{equation}
\rho_{H} = \mu_0(R_{H}H + R_{M}M)
\end{equation}

where $R_{H}$ and $R_{M}$ are the normal and anomalous Hall coefficients, $\mu_0$ is
the vacuum permeability. The first term is the ordinary Hall resistivity that
arises from the Lorentz force and the second term illustrates the anomalous Hall
contribution that originates from the intrinsic magnetisation. Figure~5c
presents $\rho_H$ versus applied field for the film CTS-650+A. $\rho_H$
initially increases with field before tending towards saturation. A weak linear
increase in $\rho_H$ at higher field emerges from the contribution of the
ordinary Hall effect. Contrary to the magnetic behavior, where higher
magnetization is observed at lower temperature, $\rho_H$ of 1~$\mu\Omega$cm is
obtained at 50~K and increases to 1.5~$\mu\Omega$cm at 300~K. The Hall conductivity 
($\sigma_{H}$) has been calculated using the formula

\begin{equation}
   \sigma_{H} = \frac{\rho_{H}}{[\rho_{xx}^{2}+\rho_{H}^{2}]} .
\end{equation}

The field dependence of the Hall conductivity $\sigma_{H}$ is plotted in
Figure~5d. $\sigma_{H}$ increases rapidly under a relatively small field due to
the anomalous contribution. This anomalous contribution has values of 284~S/cm
and 180~S/cm at 50~K and 300~K, respectively. In contrary, CTS-600 shows a Hall
conductivity of about 35~S/cm at 50~K (inset of Figure~5b). The Hall coefficient
$R_{H}$ for CTS-650+A is obtained from the slope of $\rho_{H}$ versus $H$. It is
observed that the Hall coefficient is positive for all temperatures, which
suggests that the charge carriers are of hole type, or correspondingly, the
electrons have a negative effective mass. The carrier concentration $n=e/R_{H}$
is calculated from the Hall constant $R_{H}$, where $e$ is the electron charge.
The carrier concentration is calculated to be $1.57\times10^{22}$~cm$^{-3}$ at
300~K, which corresponds to a metallic kind of behavior.

\begin{figure}[htb]
\centering
\includegraphics[angle=0,width=6.5cm,clip]{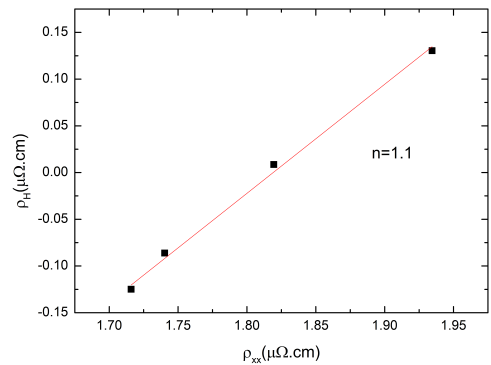}
\caption{\label{FIG 6}(Color online) 
         Anomalous Hall resistivity versus linear resistivity plotted in log$_{10}$ form
         for CTS-650+A.}
\end{figure}

Further, we discuss the physical origin of the anomalous Hall effect in CTS-
650+A. From the Berry curvature calculations, it is found that a anomalous Hall
conductivity of 100~S/cm is attained in Co$_{2}$TiSn, whereas, the measurement
gives an anomalous Hall conductivity of $\approx284$~S/cm. In order to
understand the difference, it is necessary to find out whether there is any
contribution from any extrinsic mechanisms, such as, skew scattering or side
jump effect~\cite{Smith32,Berger33}. An easier approach to explain this
mechanism is by verifying the scaling law between $\rho_{H}$ and $\rho_{xx}$ by
the formula $\rho_{H}\propto\rho_{xx}^{n}$. Here, $n$ equal to 1 and 2
correspond to the skew scattering and side jump mechanisms,
respectively~\cite{Nagaosa17}. In Figure~6, the Hall resistivity is plotted as a
function of the linear resistivity in log$_{10}$ form. The data are fitted well
by a linear dependence with a slope of $n=1.1$. This suggests that the extrinsic
effect, that is the skew scattering mechanism, may dominate in this case. It is
important to mention here that a linear conductivity $\sigma_{xx}$ of
$\approx105$~S/cm is observed in the given film. This value falls into the
category of a moderate conductivity regime, where it is considered that the
anomalous Hall effect mainly arises from the intrinsic
mechanism~\cite{Nagaosa17}. However from the current analysis presented in
Figure~6, it seems that skew scattering also contributes to the anomalous Hall
effect in the investigated film.

\section{SUMMARY}

In summary, Berry curvature calculations have been performed for Co$_2$TiSn and
an intrinsic anomalous Hall conductivity of 100~S/cm is calculated. The
anomalous Hall conductivity originates from slightly gapped nodal lines due to a
symmetry reduction induced by the magnetization. From the present analysis, it
is also found that Weyl points exist few hundreds meV above the Fermi energy.
The position of the Weyl nodes may be manipulated to appear at the Fermi energy
by doping Co$_2$TiSn to increase the valence electron concentration.  Moreover,
the valence band HAXPES spectra measured for the well-ordered Co$_2$TiSn film
suggests that the electronic structure in this film is in agreement to the
calculated density of states. Experimentally, the well-ordered film exhibits an
anomalous Hall conductivity of 284~S/cm and 180~S/cm at 50~K and 300~K,
respectively. These values are in good agreement with the theoretical
prediction. The slightly larger value of the measured anomalous Hall
conductivity is attributed to the skew scattering mechanism. The present work
emphasizes the potential of Co$_2$TiSn for future spintronics applications.

\section{ACKNOWLEDGEMENT}
The authors thank J.~K\"{u}bler for useful theoretical discussions, C.~Shekhar
for measuring transport properties, and E.~Ikenaga for experimental support
during the HAXPES measurements. The synchrotron radiation
HAXPES experiments were performed at BL47XU of Spring-8 with approval of JASRI
(Proposal No.2016B1086). This work was financially supported by the Max Planck
Society and by the ERC Advanced Grant (291472) "Idea Heusler".


\begin{thebibliography}{100}

\bibitem{Parkin1} S. S. P. Parkin, X. Jiang, C. Kaiser, A. Panchula, K. Roche, and M. Samant, Proc. IEEE {\bf91}, 661-680 (2003).

\bibitem{Pulizzi2} F. Pulizzi, Nature Mater. {\bf11}, 367 (2012).

\bibitem{FeFe2013} C.~Felser and G.~H. Fecher (Eds.). \newblock {\em Spintronics; From Materials to devices}.
                   \newblock Springer Verlag, Dordrecht, Heidelberg, New York, London, 2013.

\bibitem{Graf6} T. Graf, C. Felser, and S. S. P. Parkin, Prog. Solid State Chem. {\bf39}, 1 (2011).

\bibitem{Kubler4} J. K{\"u}bler, G. H. Fecher, and C. Felser, Phys. Rev. B. {\bf76}, 024414(2007).

\bibitem{Kandpal5} H. C. Kandpal, G. H. Fecher, and C. Felser, J. Phys. D: Appl. Phys. {\bf40}, 1507(2007).

\bibitem{Sahoo7} R. Sahoo, L. Wollmann, S. Selle, T. H{\"o}che, B. Ernst, A. Kalache, C. Shekhar, N. Kumar, S. Chadov, C. Felser, S. S. P. Parkin, and A. K. Nayak, Adv. Mater. {\bf28}, 8499 (2016).

\bibitem{Kubler8} J. K{\"u}bler, A. R. Williams, and C. B. Sommers, Phys. Rev. B. {\bf28}, 1745 (1983).

\bibitem{Galanakis9} I. Galanakis, P. H. Dederichs, and N. Papanikolaou, Phys. Rev. B. {\bf66}, 174429 (2002).

\bibitem{Sato10} J. Sato, M. Oogane, H. Naganuma, and Y. Ando, Appl. Phys. Express {\bf4}, 113005 (2011).

\bibitem{Tezuka11} N. Tezuka, N. Ikeda, F. Mitsuhashi, and S. Sugimoto, Appl. Phys. Lett. {\bf94}, 162504 (2009).

\bibitem{Liu12} H. Liu, Y. Honda, T. Taira, K. Matsuda, M. Arita, T. Uemura, and M. Yamamotoa, Appl. Phys. Lett. {\bf101}, 132418 (2012).

\bibitem{Meinert13} M. Meinert, J. Schmalhorst, H. Wulfmeier, G. Reiss, E. Arenholz, T. Graf, and C. Felser, Phys. Rev. B. {\bf83}, 064412 (2011).

\bibitem{Barth14} J. Barth, G. H. Fecher, B. Balke, S. Ouardi, T. Graf, C. Felser, A. Shkabko, A. Weidenkaff, P. Klaer, H. J. Elmers et al. Phys. Rev. B. {\bf81}, 064404 (2010).

\bibitem{Kandpal15} H. C. Kandpal, V. Ksenofontov, M. Wojcik, R. Seshadri, and C. Felser, J. Phys. D: Appl. Phys. {\bf40}, 1587 (2007).

\bibitem{Barth16} J. Barth, G. H. Fecher, B. Balke, T. Graf, A. Shkabko, A. Weidenkaff, P. Klaer, M. Kallmayer, H. J. Elmers, H. Yoshikawa, S. Ueda, K. Kobayashi, and C. Felser, Phil. Trans. R. Soc. A. {\bf369}, 3588 (2011).

\bibitem{Nagaosa17} N. Nagaosa, J. Sinova, S. Onoda, A. H. MacDonald, and N. P. Ong, Rev. Mod. Phys. {\bf82}, 1539 (2010).

\bibitem{Nayak18} A. K. Nayak, J. E. Fischer, Y. Sun, B. Yan, J. Karel, A. C. Komarek, C. Shekhar, N. Kumar, W. Schnelle, J. K{\"u}bler, C. Felser, and S. S. P. Parkin, Sci. Adv. {\bf2}, e1501870 (2016).

\bibitem{Nakatsuji19} S. Nakatsuji, N. Kiyohara, and T. Higo, Nature {\bf527}, 212-215 (2015).

\bibitem{OFF2015} S. Ouardi, G. H. Fecher, and C. Felser, J. Electr. Spectr. Rel. Phenom. {\bf190}, 249 (2013).

\bibitem{FF2013} G.~H. Fecher and C.~Felser, \newblock {\em Hard X-Ray Photoelectron Spectroscopy of New Materials for Spintronics}, chapter~11, page 243.
                \newblock Springer Verlag, Dordrecht, Heidelberg, New York, London, 2013.

\bibitem{Kubler20} J. K{\"u}bler and C. Felser, Europhys. Lett. {\bf114}, 47005 (2016).

\bibitem{Wang21} Z. Wang, M. G. Vergniory, S. Kushwaha, M. Hirschberger, E. V. Chulkov, A. Ernst, N. P. Ong, R. J. Cava, and B. A. Bernevig, Phys. Rev. Lett. {\bf117}, 236401 (2016).

\bibitem{Kubler22} J. K{\"u}bler and C. Felser, Phys. Rev. B. {\bf85}, 012405 (2012).

\bibitem{Kresse23} G. Kresse and J. Furthm{\"u}ller, Phys. Rev. B {\bf54}, 11169 (1996).

\bibitem{Kresse24} G. Kresse and J. Furthm{\"u}ller, Comput. Mater. Sci. {\bf6}, 15 (1996).

\bibitem{Perdew25} J. P. Perdew, K. Burke, and M. Ernzerhof, Phys. Rev. Lett. {\bf77}, 3865 (1996).

\bibitem{Marzari26} N. Marzari and D. Vanderbilt, Phys. Rev. B {\bf56}, 12847 (1997).

\bibitem{Souza27} I. Souza, N. Marzari ,and D. Vanderbilt, Phys. Rev. B {\bf65}, 035109 (2001).

\bibitem{Mostofi28} A. A. Mostofi, J. R. Yates, Y. S. Lee, I. Souza, D. Vanderbilt, and N. Marzari, Comput. Phys. Commun. {\bf178}, 685 (2008).

\bibitem{Chang29} G. Chang, S. Y. Xu, H. Zheng, B. Singh, C. H. Hsu, I. Belopolski, D. S. Sanchez, G. Bian, N. Alidoust, H. Lin, and M. Z. Hasan, Scientific Reports. {\bf6}, 38839 (2016).

\bibitem{Sun30} Y. Sun, Y. Zhang, C. Felser, and B. Yan, Phy. Rev. Lett. {\bf117}, 146403 (2016).

\bibitem{Sun31} Y. Sun, Y. Zhang, C. X. Liu, C. Felser, and B. Yan, arXiv:1701.09089 (2017).

\bibitem{Zegenhagen34} Detailed reviews of the method are found in Proceedings of the Workshop on Hard X-Ray Photoelectron Spectroscopy, J. Zegenhagen and C. Kunz, [Nucl. Instrum. Methods Phys. Res., Sect. A {\bf547}, No. 1 (2005).

\bibitem{Fecher35} G. H. Fecher, B. Balke, A. Gloskowskii, S. Ouardi, C. Felser, T. Ishikawa, M. Yamamoto, Y. Yamashita, H. Yoshikawa, S. Ueda, and K. Kobayashi, Appl. Phys. Lett. {\bf92}, 193513 (2008).

\bibitem{Ouardi36} S. Ouardi, G. H. Fecher, X. Kozina, G. Stryganyuk, B. Balke, C. Felser, E. Ikenaga, T. Sugiyama, N. Kawamura, M. Suzuki, and K. Kobayashi, Phys. Rev. Lett. {\bf107}, 036402 (2011).

\bibitem{Smith32} J. Smith, Physica. {\bf24}, 39 (1958).

\bibitem{Berger33} L. Berger, Phys. Rev. B. {\bf2}, 4559 (1970).


\end{thebibliography}
\end{document}